# Surface states in bulk single crystal of topological semimetal $Co_3Sn_2S_2$ towards water oxidation

Guowei Li,[1†*] Qiunan Xu,[1†] Wujun Shi,[2] Chenguang Fu,[1] Lin Jiao,[1] Machteld E. Kamminga,[3] Mingquan Yu,[4] Harun Tüysüz,[4] Nitesh Kumar,[1] Vicky Süß,[1] Rana Saha,[5] Abhay K. Srivastava,[5] Steffen Wirth,[1] Gudrun Auffermann,[1] Johannes Gooth,[1] Stuart Parkin,[5] Yan Sun,[1*] Enke Liu,[6*] and Claudia Felser[1*]

1. Max Planck Institute for Chemical Physics of Solids, 01187 Dresden, Germany
2. School of Physical Science and Technology, ShanghaiTech University, 201203 Shanghai, China
3. Zernike Institute for Advanced Materials, University of Groningen, 9747 AG Groningen, The Netherlands
4. Max Planck Institute for Coal Research, Kaiser-Wilhelm-Platz 1, 45470 Mülheim an der Ruhr, Germany
5. Max Planck Institute for Microstructure Physics, 06120 Halle, Germany
6. Institute of Physics, Chinese Academy of Sciences, 100190 Beijing, China

**Abstract**

The band inversion in topological phase matters bring exotic physical properties such as the emergence of a topologically protected surface states. They strongly influence the surface electronic structures of the investigated materials and could serve as a good platform to gain insight into the catalytic mechanism of surface reactions. Here we synthesized high-quality bulk single crystals of the topological semimetal $Co_3Sn_2S_2$. We found that at room temperature, $Co_3Sn_2S_2$ naturally hosts the band structure of a topological semimetal. This guarantees the existence of robust surface states from the Co atoms. Bulk single crystal of $Co_3Sn_2S_2$ exposes their Kagome lattice that constructed by Co atoms and have high electrical conductivity. They serves as catalytic centers for oxygen evolution process (OER), making bonding and electron transfer more efficient due to the partially filled $e_g$ orbital. The bulk single crystal exhibits outstanding OER catalytic performance, although the surface area is much smaller than that of Co-based nanostructured catalysts. Our findings emphasize the importance of tailoring topological non-trivial surface states for the rational design of high-activity electrocatalysts.

**Introduction**

Heterogeneous catalytic reactions such as the electrochemical water spitting is closely related to the surface electronic structures of the catalysts such as the surface states and surface atomic termination.(*1, 2*) The topological phase materials, with rich exotic physical properties, provides an ideal platform to explore the interplay between surface states, electron transfer, and surface catalytic reactions.(*3, 4*) Three-dimensional topological insulators have robust metallic surface states that cover the entire material. Unlike the easily destroyable surface states derived from dangling bonds, vacancies, or doping, topological surface states (TSSs) are a result of the inversion of the bulk bands at the surface. Thus, they are robust against surface modifications and defects.(*5-7*) More importantly, the electron spin is in a lock-up state with its momentum due to the spin-orbit coupling at the crystal surface. This could significantly depress backscattering and Anderson localization of conduction electrons, which is imperative for materials that always accompanied with some extent surface defects.(*8*) However, the insulating properties of topological insulators is a challenge for the electron migration when used as electrocatalysts, which will lead to lower apparent catalytic activity in comparison with that of a highly conducting catalyst.(*9*)

The OER process is a kinetically sluggish process that involves the formation of bonds and electron transfer between the catalytic sites and the adsorbates.(*10, 11*) Thus, the reaction kinetics are jointly controlled by the geometric properties (size, shape, crystallinity, etc.) and electronic structure (work functions, *d*-band center positions, spins, etc.) of the catalysts. Recently, $e_g$ orbital filling and spin states of the OER active sites has been identified to be a reasonable descriptor of catalytic activity based on the idea that the $e_g$ orbitals can form strong bonds with the oxygenated adsorbates.(*12*) It is found that depending on the spin states of the transferred electrons, either ground-state triplet oxygen molecule or hydrogen peroxide can be produced. This greatly affect the needed overpotential to drive the reaction.(*13*) Thus, it is expected excellent OER catalytic activities can be achieved by introducing elemental vacancies, applying strain, or tuning transition-metal coordination and spin states.(*14-17*) However, the strategies based on extrinsic modifications are inevitably accompanied with crystal collapse and distortion, making the exploration of their influence on catalytic activity more difficult.

Herein, taking the bulk single crystal of the topological semimetal $Co_3Sn_2S_2$ as a proof-of-concept study, we demonstrate a unique strategy to combine the advantages of a normal semimetal and a topological insulator with robust surface states, which could significantly enhance the OER kinetics. $Co_3Sn_2S_2$ was recently discovered as the first magnetic Weyl semimetal with time reversal symmetry breaking, showing a giant anomalous Hall effect in the bulk and potential topological surface states on the crystalline surface.(*18, 19*) At room temperature, the band structure of $Co_3Sn_2S_2$ naturally hosts the electronic structure of a topological semimetal. We observed high conductivity, as well as robust surface states, derived by Co atoms on the Kagome lattices and located just above the Fermi level. The $e_g$ orbital of the surface Co atoms is partially filled and points to the *p* orbital of the adsorbed hydroxide ions, thus favoring electron transfer and strengthens the bonds between the adsorbate and catalytic sites. When used as an electrocatalyst for the OER, the bulk single crystal $Co_3Sn_2S_2$ shows high activity and comparable to that of reported Co based nanostructures with a much larger surface area. The present work reveals a valuable method to develop an efficient OER electrocatalyst by manipulating the surface states and spin states.

**Results**
**Motivation**
$Co_3Sn_2S_2$ is chosen in this study because of the following reasons: 1) $Co_3Sn_2S_2$ is the first experimentally confirmed magnetic Weyl semimetal, with the existence of a Co atom-derived topologically protected surface states.(*18*) 2) Co atoms generally serve as active centers for the OER.(*20-22*) and 3) The observed high conductivity, robust surface states, as well as the magnetic Co ions in this compound indicate an intrinsic high OER activity.(*19, 23*) The crystal structure of $Co_3Sn_2S_2$ is shown in Figure 1a. It belongs to the Shandite family, exhibiting hexagonal Kagome lattices with the space group 166 (R-3*m*). Co atoms occupy Wyckoff position 3e, while S atoms are located in position 2c with *z*(S) = 0.216. There are two types of Sn atoms, which occupy positions 1a and 1b.(*23*) This arrangement can be viewed as a quasi-two-dimensional (2-D) structure stacked along the *z* direction with the Sn-[S-($Co_3Sn$)-S] layer group. The Co atoms form a Kagome lattice network with one Sn atom located at the center (Figure 1a, upper right). Thus, freshly cleaved surfaces of the measured single crystals for catalysis always exposes the (001) facet with three different terminations: six Sn atoms (1), six S atoms (2), and the Kagome lattice with six Co atoms and one Sn atom (3). The 3-dimensional (3D) Brillouin zone (BZ) and the corresponding (001) surface BZ are shown in Figure 1b. The calculation details can be seen in the Supporting Information.

To predict the room-temperature electrochemical behavior, we analyzed the electronic structure of $Co_3Sn_2S_2$ in the paramagnetic state. From the ionic picture, due to the non-closed shell configuration of the valence electrons ($3 \times Co - 3d^7 + 2 \times Sn - 5p^2 + 2 \times S - 3p^4$), a metallic nature is expected in this compound. The charge carrier density is determined to be around $1.22 \times 10^{21}$ cm$^{-3}$, showing the semi-metallic characteristics. we show the band structure in the paramagnetic state without spin-orbital coupling (SOC), as depicted in Figure 1c shows the electronic band structure without inclusion of spin orbital coupling (SOC). Owing to the crystal mirror symmetry, the band inversion induced the linear crossing near the Fermi energy can be found around the L point of the BZ, which is consistent with the reported Weyl semimetal state in ferromagnetic $Co_3Sn_2S_2$ (*19*). These nodal lines are gapped everywhere by SOC as shown in Figure 1d and Figure S1, allowing to define a $Z_2$ invariances. As given in TABLE 1, a non-zero $Z_2$ of (1; 000) can be found in $Co_3Sn_2S_2$ if the Fermi level is located within the band gap exactly.

Subsequently, we investigated the surface band structure with Sn and S terminations, respectively. Figure 1e shows the nontrivial surface states on the (001) facet for Sn termination, as calculated by using Green's function based on the tight-binding Hamiltonian. Nontrivial surface states can be observed, but are difficult to distinguish because of the weak SOC effect. With increasing strength of the SOC, the upper surface states become non-trivial here (Figure S2, surface states with 2SOC and 3SOC). Furthermore, we calculated the contribution of the Co (Figure 1f), Sn, and S atoms (Figure S3) to the nontrivial surface states (surface states with S and Sn) and found that almost all of the surface states are derived from the Co atoms. The same results were obtained when exposing the S layer (Figure S4). Further orbital analysis indicated that almost all of these surface states are almost from the Co *d* orbital. These TSSs are unoccupied and located only 0.23 eV above the Fermi level, in addition to being non-trivial and robustness against static perturbations that preserve the relevant symmetries. Finally, we calculated the decay depth of the surface states, which is estimated to be approximately 30 unit cells in the bulk. All these observations suggest that $Co_3Sn_2S_2$ is the ideal candidate to be used for exploring the electron transfer kinetics in the water oxidation process.

**Electrochemical behavior on bulk single crystal surface**
To confirm the role of *d*-derived surface states from Co atoms, high-quality bulk single crystals with the desired surface terminations are required. Here we developed a self-flux method for the synthesis of large-size $Co_3Sn_2S_2$ single crystals.(*18*) OER activities was measured in a conventional three-electrode cell containing 1 M KOH solution at a low scan rate of 5 mV s$^{-1}$ to minimize capacitive currents. A cuboid-shaped bulk single crystal was attached to a Cu wire by silver paint and used as the working electrode. Figure 2a shows the *i*R corrected linear-sweep voltammogram (LSV) curve of the bulk single crystal. When the thermodynamic OER potential ($E^0$ $H_2O/O_2$ = 1.23 V) is used as the reference, an overpotential of just 300 mV is required to reach a current density of 10 mA cm$^{-2}$. This value is close to, or even smaller than that for nanostructured electrocatalysts with a considerably larger surface area (Figure 2b), such as (Ni/Co)$_{0.85}$Se nanotube arrays (255 mV),(*24*) CoN nanowires (290 mV),(*25*) CoSn$_2$ nanocrystals (299 mV),(*26*) and NiCo metal-organic framework nanosheets (371 mV).(*27-29*) Crushing the bulk single crystal into small particles and deposited these onto Ni foam results in a further enhanced OER performance with an overpotential of 270 mV at 10 mA cm$^{-2}$. The poor activity of Ni foam suggests that the high catalytic activity of this sample originates from the Co$_3$Sn$_2$S$_2$ phase.

The catalytic kinetics for oxygen evolution is assessed by analysis of the corresponding Tafel plots. As shown in Figure 2c, the resultant Tafel slope of the Co$_3$Sn$_2$S$_2$ single crystal is only 74 mV dec$^{-1}$, which is significantly lower than that of Ni foam (190 mV dec$^{-1}$) and Co$_3$Sn$_2$S$_2$ microcrystals (95 mV dec$^{-1}$), in spite of the fact that the latter two catalysts have much smaller surface areas. This result indicates the superior reaction kinetics on the bulk single-crystal catalyst. In addition, it is interesting to observe that the linear region of the Tafel slope is much wider than for most reported studies. It is well established that the Tafel analysis is based on the Butler-Volmer equation under the assumption of constant coverage of the intermediate species. However, both surface coverage of the intermediate species and the reaction constant are strongly potential-dependent.(*30*) This well explains the rapid increase of the Tafel slope in the high-overpotential range. For Pt in 1 M KOH, this value increases from 60 to 120 mV dec$^{-1}$ when increasing the applied potential.(*31, 32*) The validity of the Butler-Volmer equation at such a large applied potential suggests a fast electron transfer kinetics on the bulk single crystal surface. Multistep chronopotentiometry measurements were performed to characterize the kinetic behavior of OH group insertion. As shown in Figure 2d, the current densities show a rapid response to the applied potential and remain stable in the following 500 s test. This suggests that the charge-transfer process is kinetically so fast, that mass transport of the active species (OH$^-$) to the crystal surface right at the moment when the potential is abruptly changed.(*33*) The 12 h durability test reveals the high stability of the microcrystal with negligible loss of the anodic current (Figure S5a). This is further confirmed by the imperceptible variation in the LSV curve after the stability test (Figure S5b).

**Phase and physical properties**

To understand the excellent OER activity in Shandite Co$_3$Sn$_2$S$_2$, the phase and physical properties of the bulk single crystal are investigated in detail. Figures 3a and S6 show the SEM image of a typical crystal for physical and electrocatalysis measurements. Bulk single crystals with dimensions of up to several centimeters can be grown, and can easily be exfoliated into lamella. The sharp and clear ordered diffraction spots from the Laue diffraction pattern confirmed the high quality of the single crystal (Figure 3a). This high quality and the purity were further confirmed by powder XRD and EDS spectra (Figure S7-8). A possible structural transition is excluded by performing single-crystal X-ray diffraction (XRD) measurements down to 100 K using a Bruker D8 Venture diffractometer. As revealed by the refinement parameters (Table S1), the R-3*m* space group is maintained throughout the measured temperature range, with only a slight thermal expansion. A shown in Figure S9 (Crystallographic Information File at 100 and 300 K can be seen in the SI), the crystal structure at 100 K is characteristic to the Shandite family; the Co atoms, which are octahedrally coordinated by four Sn and two S atoms, form a Kagome net perpendicular to the *c*-axis. However, the cobalt-centered octahedra are compressed, with significantly shorter Co-S distances (2.17 Å) than the Co- Sn (2.67 Å) distances (Table S3). For high-resolution transmission electron microscopy (TEM) observations, a thin lamella was fabricated by focused ion beam (FIB) micromachining. As shown in Figures 3b and S10, the well-defined lattice fringes with a spacing of 0.27 nm could be readily indexed to the $(10\bar{1}4)$ plane of hexagonal $Co_3Sn_2S_2$. The exposed surface is (001), as confirmed by the selected-area electron diffraction (SAED) pattern that was recorded along the [001] direction (inset in Figure 3b). The temperature ($T$) dependence of resistivity measurements was measured in four-probe configuration from 2 to 300 K along $a$ and $c$ axes, respectively. The low anisotropy of the electrical resistivity indicated that the Co $d$ electrons are itinerant. The room-temperature resistivity along the $a$ axis (exposing the (001) surface) was only 337 μΩ cm, which is much lower than that for nanostructured electrocatalysts deposited on an electrode or conductive substrates.(*16, 34*) The carrier concentration is determined to be $1.22 \times 10^{21}$ cm$^{-3}$ by Hall measurement. The combination of high conductivity and high carrier concentration in the single crystal, could improve the electrocatalytic activity significantly.(*35*) The Curie temperature ($T_C$) was determined to be 175 K from zero-field cooling (ZFC) and field cooling (FC) measurements of magnetization (Figure S11). Susceptibility follows the Curie-Weiss law and an effective moment of 0.31 $\mu_B$/Co was observed (Figure 3d). This is consistent with the results of band structure calculations and the fact that, the band dispersion near $E_F$ is mainly dominated by the Co 3$d$ orbitals and having a polarized magnetic moment of about 0.33 $\mu_B$/Co.

**Surface electronic structures**

High-resolution X-ray photoelectron spectroscopy (XPS) provides additional surface information for single-crystal catalysts (Figure S12). As shown in Figure 4a, the S $p_{3/2}$ band with a binding energy (BE) of 162.1 eV corresponds to the $S^{2-}$ configuration in $Co_3Sn_2S_2$. This value is lower than that of its counterpart $Ni_3Sn_2S_2$ (162.8 eV), which has been confirmed to adopt the electronic configuration $(Ni^0)_3(Sn^{2+})_2(S^{2-})_2$, suggesting a partial positive charge on Co atoms.(*36*) In addition, a shoulder peak with a BE of 161.4 eV is observed and can be interpreted as a surface-derived contribution.(*37*) This indicates the exposure of the $S_2$-Co-$Sn_4$ octahedra when exfoliating the bulk crystal. High-resolution Co 3$d$ spectra provide more interesting information, as shown in Figure 4b. The peak at BE = 778.3 eV can be attributed to the Co (0) states in $Co_3Sn_2S_2$, but is slightly higher than the value of metallic Co (778.1 eV),(*38*) further demonstrating the partial positive charge. A close investigation of the Co (0) 3$d_{3/2}$ peak reveals an asymmetric line shape and a small plasmonic energy loss structure, which are characteristic of a good metallic sample.(*39, 40*) This is consistent with the band structure calculations suggesting that the Fermi surface is dominated by Co 3$d$ states. The BE of 781.1 eV can be ascribed to $Co^{2+}$, which is a result of surface oxidation or loss of coordination such as S vacancies. The clearly distinguished, broad peak centered at 785.2 eV is a satellite structure for $Co^{\delta+}$ and $Co^{2+}$. (*39*) The Sn 3$d$ spectra clearly indicate that the main peak is derived from the $Sn^{2+}$ states (Figure S13). All these results suggest that the bulk single crystal $Co_3Sn_2S_2$ has a different electronic configuration from that of $Ni_3Sn_2S_2$, more likely to be $(Co^{\delta+})_3(Sn^{2-\gamma})_2(S^{2-})_2$, with the Co atoms being partially positively charged and Sn atoms having an average valence below 2. For direct determination of the surface atomic termination, a bulk $Co_3Sn_2S_2$ single crystal was cleaved under ultrahigh vacuum conditions to expose the (001) surface. Scanning tunneling microscope (STM) images were collected in situ at 2 K and are shown in Figure 4c. The cleaved surface exhibits a typical Kagome-lattice atomic structure, indicating a Co layer at the surface. Such a cleave is energetically favorable because it exposes the smallest surface areas, and the S-Sn bond can be easily broken because of the large bonding distance (2.86 Å) (Table S3).

**Discussion**

The high-quality bulk single crystal with well-defined surface termination provides an ideal platform for understanding the surface catalytic process. Although the details of the mechanism of the O=O bond formation remain unresolved, there is a consensus that the initial discharge of hydroxide ions at a catalytically active center is the initial step. For transition metals such as Co, Ni, and Fe, the crystal field formed by different types of coordination results in distinct spin states and $e_g$ filling. In octahedrally coordinated systems, the $e_g$ orbital has a large overlap with the oxygen related adsorbates, making electron transfer between the active sites and adsorbates more favorable. This results the $e_g$ filling of surface active centers as an activity descriptor for OER and ORR.(*12, 41*) For our $Co_3Sn_2S_2$ single crystals, the Co atoms in the bulk are octahedrally coordinated and the d orbitals are split into three-fold degenerate $t_{2g}$ states lying lower in energy, and two-fold $e_g$ states at higher energy (Figure 4d and Figure S14). The intermediate spin states of Co with a magnetic moment of 0.31 $\mu_B$/Co. indicates that the $e_g$ orbitals are half-filled. However, for the surface Co atoms, the cleaving between S and Sn leads to loss of coordination and further breaking of the degeneracy of the $t_{2g}$ and $e_g$ orbitals, as illustrated in Figure 4d.(*42, 43*) The half-filled $d_{z^2}$ orbital point toward the *p* orbital of the adsorbed hydroxide ions resulting from the interaction with the bridging $O^{2-}$ via π-donation. This gives rise to the formation of σ-bonding between the Co atom and the surface $OH^-$ adsorbates and favors electron transfer between them.(*12, 14, 27*) In addition, loss of coordination for the surface Co octahedral leads to the formation of a highly distorted five-fold coordinated square pyramids. The newly open coordination sites make the uptake of $OH^-$ more favorable.(*15*) More importantly, our theoretical investigations and previous ARPES measurements confirms that the surface states derived by Co are topologically protected. Different from the spin degeneracy caused by fabricating elemental vacancies, defects, or doping, which are easily destroyed by breaking the surface symmetry, the unoccupied TSSs of Co provide robust active sites for oxygen evolution. Based on the reaction mechanism, the kinetics of $OH^-$ bonding *via* the surface Co atoms are determined by theoretical studies. Indeed, as shown in Figure S15, $OH^-$ binds to Co atoms and is located above the center of the three Co atoms. The bonding distance is determined to be 2.00 Å, which is shorter than the original Co-Sn bonding distance of 2.67 Å. The iso-surface plot of the transferred charge distribution for the slab of the Co octahedral and $OH^-$ adsorbate is shown in Figure 4e. It can be seen clearly that the interfacial charge distributions on the Co atom are dumbbell-like, indicating a $3d_{z^2}$ orbital order,(*43*) while the charge distribution on Sn is strictly spherical. The total charge on $Co^{\delta+}$ is calculated to be 8.75, while this value for $OH^-$ is 7.67, vividly confirming electron transfer through the Co-O bonding. Very interestingly, we also observed the similar electron transfer behavior on other adsorbate such as hydrogen (Figure S16). As two of the most important reaction intermediates of many surface reactions such as hydrogen evolution, hydrogen reduction, oxygen reduction, and $H_2O_2$ electrochemical synthesis, ore results indicate that one can efficiently control the surface reactions by carefully tailoring the robust surface states of topological phase catalysts.

In conclusion, we synthesized high-quality $Co_3Sn_2S_2$ bulk single crystals with well-defined atomic surface termination. Our magnetic, electrical resistivity, electrochemical measurements, XPS, STM, and DFT calculations uncovered the local spin structure and ligand environment of the surface Co atoms. The results provide new insights into the understanding of the surface water oxidation process. First, the non-trivial surface states derived by Co atoms are robust to surface distortion and modification. More importantly, they are located just above the Fermi energy and can accept electrons from the adsorbates. In addition, loss of coordination of the Co atoms provides new open sites for the uptake of $OH^-$ species. This facilitates the electron transfer through coupling between the $e_g$ orbital of Co and the O-$p_\sigma$ orbital of $OH^-$. The present study provides a promising strategy to create highly efficient and robust catalysts by utilizing the surface states around the Fermi energy.

**Materials and Methods**
**Materials synthesis.**
$Co_3Sn_2S_2$ polycrystals were obtained by mixing high-purity elements with desired stoichiometry. The polycrystalline samples of $Co_3Sn_2S_2$ were sealed in a quartz tube with some iodine under a partial Argon pressure. The samples were heated to 1000 °C over 6 hours and kept for 24 hours before being slowly cooled to 600 °C over 7 days. Large single-crystalline ingots up to centimeter size were obtained at the bottom of the quartz tubes.

**Characterization.**
Single crystal X-ray diffraction (XRD) measurements were performed using a Bruker D8 Venture diffractometer equipped with a Triumph monochromator and a Photon100 area detector, operating with Mo Kα radiation. High resolution transmission electron microscopy (HRTEM) was performed on a large lamella fabricated by FIB. The longitudinal electrical resistance measurement was conducted using a standard four-probe method with the AC transport option in a PPMS system. XPS spectra were carried on a UHV surface analysis system equipped with a Scienta-200 hemispherical analyzer. The base pressure of the sample analysis chamber is $2 \times 10^{-10}$ mbar. Magnetization measurements were carried out on oriented crystals with the magnetic field applied along the *a*-axes on the vibrating sample magnetometer (MPMS 3, Quantum Design). For STM, $Co_3Sn_2S_2$ single crystals were cleaved in situ at $T < 20$ K to expose a (001) surface. After cleaving, the samples are quickly transferred to the STM head and kept in ultra-high vacuum ($p < 3 \times 10^{-9}$ Pa) and at low temperature ($T = 2$ K). The tunneling spectra were measured using tungsten tips acquired by the standard lock-in technique.

**Electrocatalytic characterization.**
The assessment of OER activities were performed on the Autolab PGSTAT302N with an impedance module electrochemistry workstation. Ag/AgCl (3 M KCl) electrode was used as the reference electrode, and a graphite rod was used as the counter electrode. The bulk $Co_3Sn_2S_2$ single crystal was attached to a Cu wire with silver paint and served as working electrode and catalyst. The Linear sweep voltammograms were recorded with a scan rate of 5 mV / s. The electrochemical impedance spectroscopy was carried out with a 10 mV AC potential from 20 kHz to 0.01 Hz for correcting the polarization curves. All potentials were referenced to a reverse hydrogen electrode (RHE).

**Acknowledgements**

The authors gratefully acknowledge Dr. J. Wu for the XPS measurement. M. Y thanks the supporting from IMPRS-RECHARGE. **Funding:** This work was financially supported by the European Research Council (ERC Advanced Grant No. 291472 'Idea Heusler') and ERC Advanced Grant (No. 742068). TOPMAT'. M. E. K. acknowledges the Netherlands Organization for Scientific Research (NWO) (No. 022.005.006). E.L. acknowledges the National Key R&D Program of China (No. 2017YFA0206303) and the National Natural Science Foundation of China (No. 51722106). The calculations were carried out at the HPC Platform of Shanghaitech University.

| TRIM points | Γ (0,0,0) | L (0.5,0,0) ×3 | F (0.5,0.5,0) ×3 | T (0.5,0.5,0.5) |
|---|---|---|---|---|
| parity | - | - | + | - |

**Table 1. $Z_2$ numbers (1; 000) of $Co_3Sn_2S_2$ crystal.** The product of parity of occupied bands at each time reversal invariant momenta (TRIM) points.

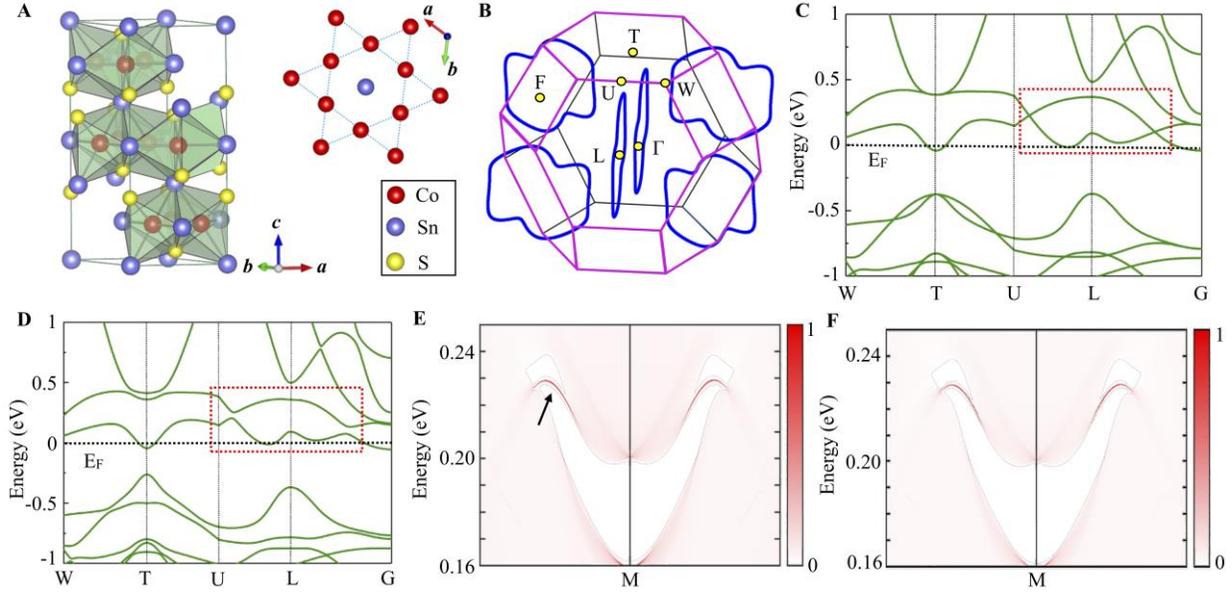

**Fig. 1. Crystal and band structure of bulk single crystal Co$_3$Sn$_2$S$_2$. a.** Crystal structure of Co$_3$Sn$_2$S$_2$ obtained from single crystal XRD and the Kagome lattice structure constructed by Co atoms in the *a-b* plane. **b.** The 3 D Brillouin zone projected in the (001) direction. Three pairs of nodal lines are shown in the first Brillouin zone. **c.** Band structure of Co$_3$Sn$_2$S$_2$ in a paramagnetic state without the consideration of SOC effect. The band linear crossing near the Fermi energy can be found around the point L. **d.** Band structure of Co$_3$Sn$_2$S$_2$ with the inclusion of SOC effect. The band linear crossing is open, resulting in the band gap. **e.** The nontrivial surface states on (001) facet of Co$_3$Sn$_2$S$_2$ crystal with Sn termination, which is not fully occupied and located just above the Fermi level. **f.** The contribution of Co atoms to the nontrivial surface states shown in **e**. Nearly all the states originate from the surface Co Kagome layer.

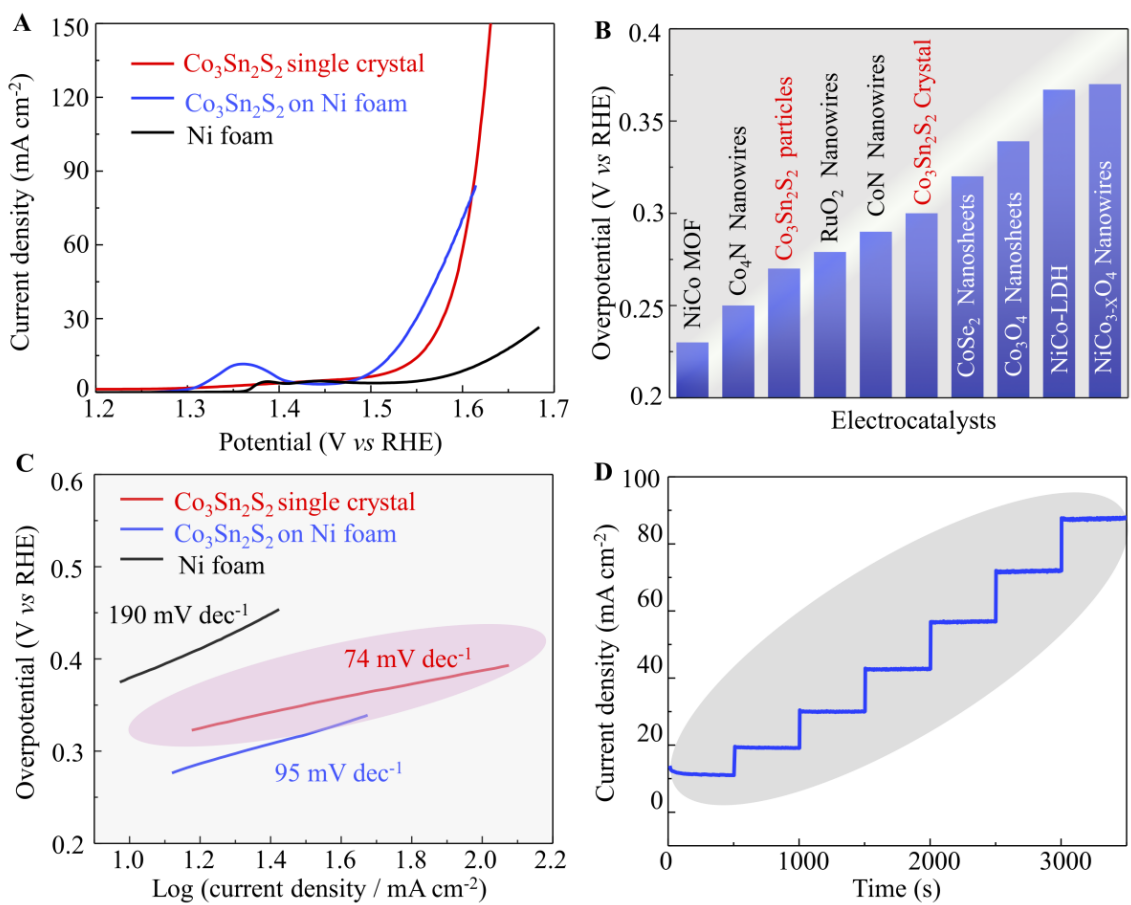

**Fig. 2. Electrochemical performance of Co₃Sn₂S₂ single crystal catalyst. a.** OER polarization curves for Ni foam, Co₃Sn₂S₂ single crystal, and Co₃Sn₂S₂ micro-powder crushed from the single crystal. **b.** Overpotential of Co₃Sn₂S₂ single crystal catalyst at 10 mA cm⁻² compared with some recently reported results for OER electrocatalysts. **c.** Tafel plot of Ni foam, Co₃Sn₂S₂ single crystal, and Co₃Sn₂S₂ micro-powder Koutecky–Levich plots in O₂-saturated 1M KOH solution. The wide linear regime indicates the excellent electron transfer kinetics even at large overpotential. **d.** Multi-current process with the current density increased from 10 to 85 mA cm⁻² without iR correction.

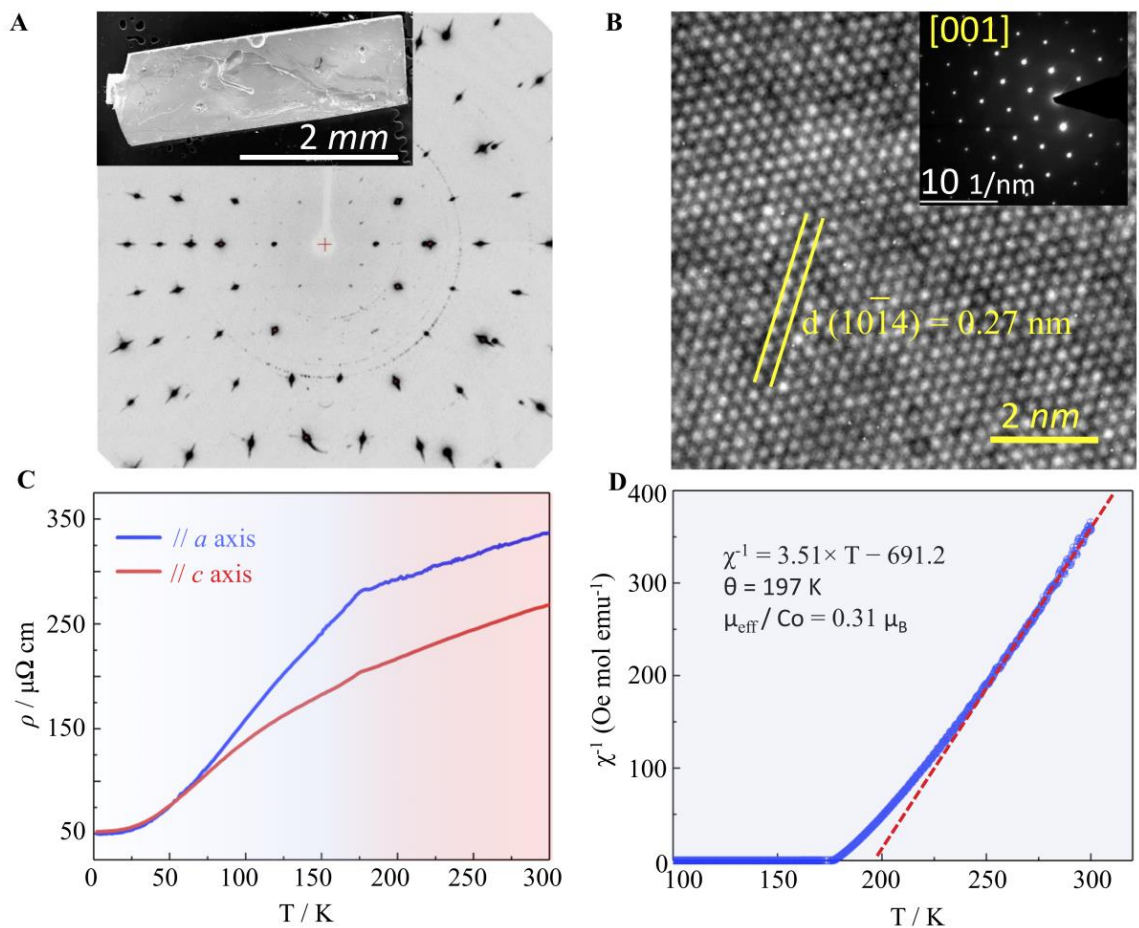

**Fig. 3. Phase structure and physical properties of Co$_3$Sn$_2$S$_2$ single crystal catalyst. a.** Single-crystal XRD pattern of Co$_3$Sn$_2$S$_2$. The pattern was recorded by rocking by 32° about the *b*-axis of the rhombohedral cell. The high quality of the crystal is proven by the clear and sharp diffraction spots. The faint rings may be attributed to distortions and contaminations on the crystal surface. A typical SEM image of the single crystal is shown in the upper-left corner. **b.** HRTEM image of the Co$_3$Sn$_2$S$_2$ single crystal prepared using the Focused Ion Beam technique (FIB) and the selected area diffraction (SAED) pattern recorded along the [001] crystal orientation. **c.** Temperature dependence of electric resistivity of Co$_3$Sn$_2$S$_2$ single crystal in zero field. The current was applied along the *a* and *c* axis. **d.** Reciprocal susceptibility as a function of temperature. The magnetic moments are derived from Co atoms in the Kagome lattice. Employing Curie Weiss law, an effective Bohr magneton $\mu_{eff}$ of 0.31 $\mu_B$/Co is obtained.

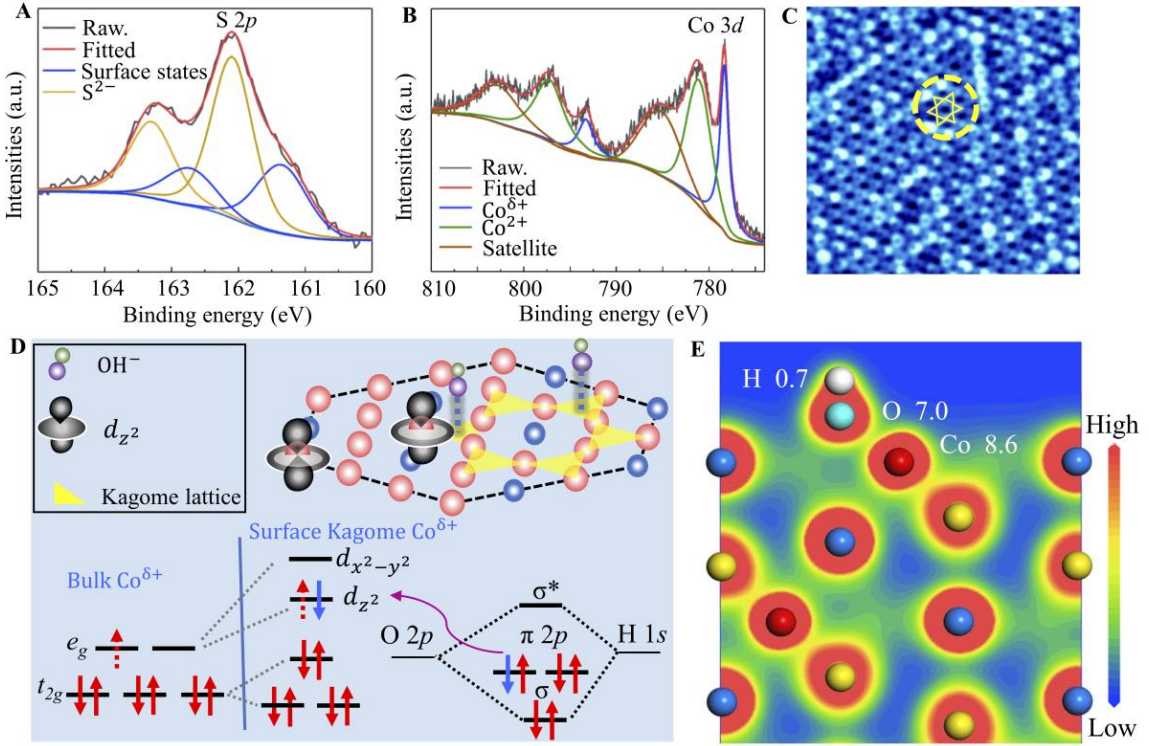

**Fig. 4. Surface structure and OER mechanism. a.** Detailed XPS analysis of the prepared single crystal. High resolution XPS spectra for **a.** S 2*p* and **b.** Co 3*d*. **c.** STM topography of a cleaved $Co_3Sn_2S_2$ single crystal thin flake showing an area of $8 \times 8$ nm$^2$. The Kagome lattice is highlighted by yellow lines in the circle. **d.** Schematic representation of the favored OH uptake with the Co 3*d* orbitals. The exfoliation between the S-Sn plane break the octahedral symmetry of the surface Co atoms in the Kagome lattice (highlighted by yellow triangle). The empty $3d_{z^2}$ orbital points to the *p* orbital of the OH group, resulting in a strong bonding between them. **e.** Contour plots of the total charge distribution of $Co_3Sn_2S_2$ single crystal with one OH group bonded to the Co atoms. Electronic charges are distributed in the vicinity of Sn atoms. However, for Co atoms, one can see the electron transfer through the Co-O bonding.

16